# On approximations via convolution-defined mixture models

Hien D. Nguyen and Geoffrey J. McLachlan

## Abstract

An often-cited fact regarding mixing or mixture distributions is that their density functions are able to approximate the density function of any unknown distribution to arbitrary degrees of accuracy, provided that the mixing or mixture distribution is sufficiently complex. This fact is often not made concrete. We investigate and review theorems that provide approximation bounds for mixing distributions. Connections between the approximation bounds of mixing distributions and estimation bounds for the maximum likelihood estimator of finite mixtures of location-scale distributions are reviewed.

## Index Terms

Mixing distributions; Finite mixture models; convolutions; Kullback-Leibler divergence; Maximum likelihood estimators.

## I. INTRODUCTION

Mixing distributions and finite mixture models are important classes of probability models that have found use in many areas of application, such as in artificial intelligence, machine learning, pattern recognition, statistics, and beyond. Mixing distributions provide probability models with probability density functions (PDFs) of the form $f(\boldsymbol{x}) = \int_{\mathbb{X}} f(\boldsymbol{x}; \boldsymbol{\theta}) \, \mathrm{d}\Pi(\boldsymbol{\theta})$, where $f(\cdot; \boldsymbol{\theta})$ is a PDF (with respect to a random variable $\boldsymbol{X} \in \mathbb{X}$) that is dependent on some parameter $\boldsymbol{\theta} \in \Theta \subseteq \mathbb{R}^d$ with distribution function (DF) $\Pi$. Here, $\mathbb{X} \subseteq \mathbb{R}^p$ is the support of the PDFs $f$ and $f(\cdot; \boldsymbol{\theta})$, where $\mathbb{X}$ is not functionally dependent on $\boldsymbol{\theta}$. Notice that the mixing distributions contain the finite mixture models by setting the DF to $\Pi(\boldsymbol{\theta}) = \sum_{i=1}^{n} \pi_i \delta(\boldsymbol{\theta} - \boldsymbol{\theta}_i)$, for $n \in \mathbb{N}$, where $\delta$ is the Dirac delta function (cf. [1]), $\pi_i \geq 0$, $\sum_{i=1}^{n} \pi_i = 1$, $\boldsymbol{\theta}_i \in \mathbb{X}$ (for each $i \in [n] = \{1, \ldots, n\}$), and $\mathbb{N}$ is the natural numbers (zero exclusive); see [2, Sec. 1.2.2] and [3, Sec. 1.12] for descriptions and references regarding mixing distributions.

The appeal of finite mixture models largely comes from their flexibility of representation. The folk theorem regarding mixture models generally states that a mixture model can approximate any distribution to a sufficient level of accuracy, provided that the number of mixture components is sufficiently large. Example statements of the folk theorem include: "provided the number of component densities is not bounded above, certain forms of

Hien Nguyen is with the Department of Mathematics and Statistics, La Trobe University, Bundoora, Melbourne, Australia 3086. Geoffrey McLachlan is with the School of Mathematics and Physics, The University of Queensland, St. Lucia, Brisbane, Australia 4075. *Corresponding Author: Hien Nguyen (Email: h.nguyen5@latrobe.edu.au).

mixture can be used to provide arbitrarily close approximations to a given probability distribution" [4, p. 50], "any continuous distribution can be approximated arbitrarily well by a finite mixture of normal densities with common variance (or covariance matrix in the multivariate case)" [3, p. 176], "there is an obvious sense in which the mixture of normals approach, given enough components, can approximate any multivariate density" [5, p. 5], "the [mixture] model forms can fit any distribution and significantly increase model fit" [6, p. 173], and "a mixture model can approximate almost any distribution" [7, p. 500]. From the examples, we see that statements regarding the flexibility of mixture models are generally left technically vague and unclear.

Let $d_{\mathbb{X}}^{\text{TV}}(f,g) = \frac{1}{2}\|f-g\|_{\mathbb{X},1}$ be the total-variation distance, where $\|f\|_{\mathbb{X},q} = \left[\int_{\mathbb{X}}|f(\boldsymbol{x})|^q\,\mathrm{d}\boldsymbol{x}\right]^{1/q}$ for $q \in [1,\infty]$ is the $\mathcal{L}_q$-norm over support $\mathbb{X}$. Here, $f(\boldsymbol{x})$ and $g(\boldsymbol{x})$ are functions over the support $\mathbb{X} \subseteq \mathbb{R}^d$. We also let $\|f\|_{\mathbb{X},\infty} = \sup_{\boldsymbol{x}\in\mathbb{X}}|f(\boldsymbol{x})|$. Further define the location-scale family of PDFs over $\mathbb{R}$ as

$$\mathcal{F}^1 = \left\{ f : \int_{\mathbb{R}} \frac{1}{\sigma} f\left(\frac{x-\mu}{\sigma}\right)\mathrm{d}x = 1,\text{ for all } \mu \in \mathbb{R} \text{ and } \sigma \in (0,\infty) \right\}.$$

Define $x_i$ to be the $i$th element of $\boldsymbol{x}$, for $i \in [p]$. Historical technical justifications for the folk theorem include [8] and [9]. Within more recent literature, it is difficult to obtain clear technical statements of such results. Among the only references that we could find is the exposition of [10, Sec. 33.1], and in particular, the following theorem statement.

**Theorem 1** (DasGupta 2008, Thm. 33.1). *Let $f$ be a probability density function over $\mathbb{R}^p$ for $p \in \mathbb{N}$. If $\mathcal{F}_g^2$ is the class of mixtures of $g \in \mathcal{F}^1$:*

$$\mathcal{F}_g^2 = \left\{ f^* : f^*(\boldsymbol{x}) = \int_0^\infty \int_{\mathbb{R}^p} \frac{1}{\sigma^d} \prod_{i=1}^p g\left(\frac{x_i - \mu_i}{\sigma}\right) d\Pi_\mu(\boldsymbol{\mu})\, d\Pi_\sigma(\sigma) \right\}$$

*Then, given any $\epsilon > 0$, there exists a $f^* \in \mathcal{F}^2$ such that $d_{\mathbb{R}^d}^{TV}(f,f^*) < \epsilon$, where $\Pi_\mu(\boldsymbol{\mu})$ and $\Pi_\sigma(\boldsymbol{\sigma})$ are DFs over $\mathbb{R}^d$ and $(0,\infty)$, respectively.*

Upon inspection, Theorem 1 states that the class of marginally-independent location-scale mixing distributions have PDFs that can approximate any other PDF arbitrarily well, with respect to the total-variation distance. Unfortunately, the proof of Theorem 1 is not provided in [10]. Remarks regarding the theorem relegate the proof to an unknown location in [11], which makes it difficult to investigate the structure and nature of the theorem.

The lack of transparency from the cited text has lead us to investigate the nature of the presented theorem. The outcome of our investigation is the collection of the various proofs and technical results that are reviewed in this article. The contents of our review are as follows.

Firstly, we investigate the proofs from [11] and consider alternative versions of Theorem 1 that provide more insight into the structure of the result. For example, we present an alternative to Theorem 1, whereupon only a mixture over the location parameter is required. That is, no integration over the scale parameter element is needed, as in $\mathcal{F}_g^2$. Furthermore, we state a uniform approximation alternative to Theorem 1 that is applicable to the approximation of target PDFs over compact sets. Rates of convergence are also obtainable if we make a Lipschitz assumption on the target PDF.

In addition to the presentation of Theorem 1 and its variants, we also review the relationship between the mixing distributions results and the approximation bounds of [12]. Via the approximation and estimation bounding results for the maximum likelihood estimator (MLE) from [13] and [14], we further present results for bounding Kullback-Leibler errors [KL; [15]] of the MLE for finite mixtures of location-scale PDFs.

The article proceeds as follows. In Section 2, we discuss Theorem 1 and its variants. The relationship between the mixing distributions approximation results and the results of [12], [13], and [14] are then presented in Sections 3, 4, and 5, respectively.

## II. MIXING DISTRIBUTION APPROXIMATION THEOREMS

Let $\mathcal{L}_q(\mathbb{X})$ be the space of functions having the property $\|f\|_{\mathbb{X},q} < \infty$, with support $\mathbb{X}$, and which map to $\mathbb{R}$. Further, we define the convolution between $f \in \mathcal{L}_q(\mathbb{X})$ and $g \in \mathcal{L}_r(\mathbb{X})$ as

$$(f * g)(\boldsymbol{x}) = \int_{\mathbb{X}} f(\boldsymbol{y}) g(\boldsymbol{x} - \boldsymbol{y}) \, \mathrm{d}\boldsymbol{y}, \tag{1}$$

where (1) exists and is measurable for specific cases, due to results such as the following from [16, Sec. 9.3].

**Theorem 2** (Makarov and Podkorytov, 2013, Sec. 9.3.1-2). *Let $f \in \mathcal{L}_q(\mathbb{R}^p)$ and $g \in \mathcal{L}_r(\mathbb{R}^p)$, for $q, r \in [1, \infty]$. We have the following results:*

(i) *if $q = 1$, then $f * g$ exists and $\|f * g\|_{\mathbb{R}^p, r} \leq \|f\|_{\mathbb{R}^p, 1} \|g\|_{\mathbb{R}^p, r}$.*

(ii) *if $1/q + 1/r = 1$, then $f * g$ exists and $\|f * g\|_{\mathbb{R}^p, \infty} \leq \|f\|_{\mathbb{R}^p, q} \|g\|_{\mathbb{R}^p, r}$.*

*Remark* 3. When $q = r = 1$, not only do we have inequality (i) of Theorem 2, but also that

$$\begin{aligned}
\int_{\mathbb{R}^p} (f * g)(\boldsymbol{x}) \mathrm{d}\boldsymbol{x} &= \int_{\mathbb{R}^p} f(\boldsymbol{y}) \int_{\mathbb{R}^p} g(\boldsymbol{x} - \boldsymbol{y}) \, \mathrm{d}\boldsymbol{x} \mathrm{d}\boldsymbol{y} \\
&= \int_{\mathbb{R}^p} f(\boldsymbol{x}) \, \mathrm{d}\boldsymbol{x} \int_{\mathbb{R}^p} g(\boldsymbol{x}) \, \mathrm{d}\boldsymbol{x} < \infty.
\end{aligned}$$

Moreover, this implies that if $f$ and $g$ are PDFs over $\mathbb{R}^p$, then $f * g$ is also a PDF over $\mathbb{R}^p$.

Let a function $\alpha_k \in \mathcal{L}_1(\mathbb{R}^p)$ for $k \in \mathbb{R}^+$ be called an approximate identity in $\mathbb{R}^p$ if there exists a $k^* \in [0, \infty]$ such that (i) $\alpha_k \geq 0$, (ii) $\int_{\mathbb{R}^p} \alpha_k(\boldsymbol{x}) \, \mathrm{d}\boldsymbol{x} = 1$, and (iii) $\int_{\|\boldsymbol{x}\|_1 > \delta} \alpha_k(\boldsymbol{x}) \, \mathrm{d}\boldsymbol{x} \to 0$ as $k \to k^*$, for every $\delta > 0$ [cf. [16, Sec. 7.6.1]]. Here, $\|\boldsymbol{x}\|_q = \left(\sum_{i=1}^{p} |x_i|^q\right)^{1/q}$ is the $l_q$-vector norm. The following result of [11] provides a useful generative method for constructing approximate identities.

**Lemma 4** (Cheney and Light, 2000, Ch. 20, Thm. 4). *Let $\alpha \in \mathcal{L}_1(\mathbb{R}^p)$ and let $k \in \mathbb{N}$. If $\alpha \in \mathcal{F}^3$, where*

$$\mathcal{F}^3 = \left\{ f \in \mathcal{L}_1(\mathbb{R}^p) : f(\boldsymbol{x}) = \prod_{i=1}^{p} g(x_i), \int_{\mathbb{R}} g(x) \, dx = 1, \text{ and } g(x) \geq 0 \text{ for all } x \in \mathbb{R} \right\},$$

*then the dilations $\alpha_k(\boldsymbol{x}) = k^p \alpha(k\boldsymbol{x})$ is an approximate identity, with $k^* = \infty$.*

W may call $\mathcal{F}^3$ the class of marginally-independent scaled density functions. With an ability to construct approximate identities, the following theorem from [16] provides a powerful means to construct approximations for any function over $\mathbb{R}^p$. The corollary to the result provides a statistical interpretation.

**Theorem 5** (Makarov and Podkorytov, 2013, Sec. 9.3.3). *Let $\alpha_k$ be an approximate identity in $\mathbb{R}^p$ for some $k^* \in [0, \infty]$. If $f \in \mathcal{L}_q(\mathbb{R}^p)$ for $q \in [1, \infty)$, then $\|f * \alpha_k - f\|_{\mathbb{R}^p, q} \to 0$ as $k \to k^*$.*

**Corollary 6.** *Let $f$ be a PDF in $\mathcal{L}_q(\mathbb{R}^p)$, for $q \in [1, \infty)$. If $g \in \mathcal{F}^3$, then for any $\epsilon > 0$, there exists a PDF $f^*$ in*

$$\mathcal{F}_g^4 = \left\{ f^* : f^*(\boldsymbol{x}) = \int_{\mathbb{R}^p} k^p g(k\boldsymbol{x} - k\boldsymbol{m}) f(\boldsymbol{m}) d\boldsymbol{m}, \ k \in \mathbb{N} \right\}$$

*such that $\|f - f^*\|_{\mathbb{R}^p, q} < \epsilon$, for any $q \in [1, \infty)$.*

*Proof:* From Lemma 4 and Theorem 5, for any $g \in \mathcal{F}^3$ and $f \in \mathcal{L}_q(\mathbb{R}^p)$, we have $\|f * [k^p g(k \times \cdot)] - f\|_{\mathbb{R}^p, q} \to 0$, where the convolution $f * [k^p g(k \times \cdot)] = \int_{\mathbb{R}^p} k^p g(k\boldsymbol{x} - k\boldsymbol{m}) f(\boldsymbol{m}) d\boldsymbol{m}$. By the definition of convergence, we have the fact that for every $\epsilon > 0$, there exists some $K$ such that for all $k > K$, $\|f * [k^p g(k \times \cdot)] - f\|_{\mathbb{R}^p, q} < \epsilon$. Putting the convolutions $f * [k^p g(k \times \cdot)]$ for all $k \in \mathbb{N}$ into $\mathcal{F}_g^4$ provides the desired convergence result. Lastly, $f^*$ is a PDF via Remark 3. ∎

*Remark* 7. Corollary 6 improves upon Theorem 1 in several ways. Firstly, the total variation bound is replaced by the stronger $\mathcal{L}_q$-norm result. Secondly, mixing only occurs over the mean parameter element $\boldsymbol{m}$, via the PDF $d\Pi_m(\boldsymbol{m})/d\boldsymbol{m} = f(\boldsymbol{m})$, and not over the scaling parameter element $k$, which can be taken as a constant value. That is, we only require that the class $\mathcal{F}_g^4$ be mixing distributions over the location parameter element of $g$, where $\Pi_m$ is the DF that is determined by the density being approximated and the scale parameter element is picked to be some fixed value $k \in \mathbb{N}$. Lastly, we note that Theorem 1 can simply be obtained as the $q = 1$ case of Corollary 6 by setting $\sigma = 1/k$ and $\boldsymbol{\mu} = \boldsymbol{m}/k$.

Notice that Theorem 5 cannot be used to provide $\mathcal{L}_\infty$-norm approximation results. Let $\mathcal{C}(\mathbb{X})$ be the class of continuous functions over the set $\mathbb{X}$. If one assumes that the target PDF $f$ is bounded and belongs to $\mathcal{C}(\mathbb{R}^p)$, then a uniform approximation alternative to Theorem 5 is possible for compact subsets of $\mathbb{R}^p$.

**Theorem 8** (Cheney and Light, 2000, Ch. 20, Thm. 2). *Let $\alpha_k$ be an approximate identity in $\mathbb{R}^p$ for some $k^* \in [0, \infty]$. If $f$ is a bounded function in $\mathcal{C}(\mathbb{R}^p)$, then $\|f * \alpha_k - f\|_{\mathbb{K}, \infty} \to 0$ as $k \to k^*$, for all compact $\mathbb{K} \subset \mathbb{R}^p$.*

We note in passing that Theorem 8 can be used to prove density results for finite mixture models, such as that of [10, Thm. 33.2]. For further details, see [11, Thm. 5] and [17]. Let $\text{Lip}_a(\mathbb{X})$ be the class of Lipschitz functions $f$, where $|f(\boldsymbol{x}) - f(\boldsymbol{y})| \leq C \|\boldsymbol{x} - \boldsymbol{y}\|_\infty^a$, for some $a, C \in [0, \infty)$, where $\boldsymbol{x}, \boldsymbol{y} \in \mathbb{X}$. If one assumes that the target PDF is in $\text{Lip}_a(\mathbb{X})$ for some $a \in (0, 1]$, then the following approximation rate result is available.

**Theorem 9** (Cheney and Light, 2000, Ch. 21, Thm. 1). *Let $\alpha_k$ be an approximate identity in $\mathbb{R}^p$ for some $k^* \in [0, \infty]$ with the additional property that $\int_{\mathbb{R}^p} \|\boldsymbol{x}\|_1^a \alpha_1(\boldsymbol{x}) d\boldsymbol{x} < \infty$ for some $a \in (0, 1]$. If $f \in \text{Lip}_a(\mathbb{R}^p)$, then there exists a constant $A > 0$ such that $\|f * \alpha_k - f\|_{\mathbb{R}^p, \infty} \leq A/k^a$ for $k \in \mathbb{N}$.*

**Example 10.** Let $\alpha \in \mathcal{F}^3$ be generated by taking the marginal location-scale density $g = \phi$, where $\phi$ is the standard normal PDF. The condition $\int_{\mathbb{R}^p} \|\boldsymbol{x}\|_1^a \alpha_1(\boldsymbol{x}) d\boldsymbol{x} < \infty$ is satisfied for $a = 1$ since the multivariate normal distribution has all of its polynomial moments; see for example [18].

5**Corollary 11.** *Let $f \in Lip_1(\mathbb{R}^d)$ be a PDF. If $f^*(\boldsymbol{x}) = \int_{\mathbb{R}^d} k^p \prod_{i=1}^{p} \phi(kx_i - km_i) f(\boldsymbol{m}) \, d\boldsymbol{m}$, then $\|f - f^*\|_{\mathbb{R}^p, \infty} \leq A/k$ for $k \in \mathbb{N}$ and some constant $A > 0$.*

Thus, the mixing distribution generated via marginally-independent normal PDFs convergences uniformly for target PDFs $f \in \text{Lip}_1(\mathbb{R}^p)$, at a rate of $1/k$.

### III. BOUNDING OF KULLBACK-LEIBLER DIVERGENCE VIA RESULTS FROM ZEEVI AND MEIR (1997)

Let $\mathbb{K} \subset \mathbb{R}^p$ be a compact subset and let

$$\mathcal{F}^5_{\mathbb{K},\beta} = \left\{ f : \int_{\mathbb{K}} f(\boldsymbol{x}) \, d\boldsymbol{x} = 1 \text{ and } f(\boldsymbol{x}) \geq \beta > 0, \text{ for all } \boldsymbol{x} \in \mathbb{K} \right\}$$

be the class of lower-bounded target PDFs over $\mathbb{K}$. In [12], the approximation errors for finite mixtures of marginally-independent PDFs are studied in the context of approximating functions in $\mathcal{F}^5_{\mathbb{K},\beta}$.

*Remark* 12. The use of finite mixtures of marginally-independent PDFs is implicit in [12] as they report on approximation via product kernels of radial basis functions. The products of kernels is equivalent to taking products over marginally-independent densities to yield a joint density. Univariate radial basis functions that are positive and integrate to unity are symmetric PDFs in one dimension. Thus, the product of univariate radial basis functions that generate PDFs correspond to a subclass of $\mathcal{F}^3$; see [19] regarding radial basis functions.

Let the KL divergence between two PDFs $f, g \in \mathcal{L}_1(\mathbb{X})$ be defined as

$$d^{\text{KL}}_{\mathbb{X}}(f, g) = \int_{\mathbb{X}} f(\boldsymbol{x}) \log\left[\frac{f(\boldsymbol{x})}{g(\boldsymbol{x})}\right] d\boldsymbol{x}.$$

The KL divergence between $f$ and $g$ is difficult to work with as it is not a distance function. That is, it is asymmetric and it does not obey the triangle inequality. As such, bounding the KL divergence by a distance function provides a useful means of manipulation and control. The following useful result is obtained by [12].

**Lemma 13** (Zeevi and Meir, 1997, Lemma 3). *If $f, g \in \mathcal{F}^5_{\mathbb{K},\beta}$, then $d^{\text{KL}}_{\mathbb{K}}(f, g) \leq \beta^{-1} \|f - g\|^2_{\mathbb{K},2}$.*

Let $\mathcal{F}^6_g = \{k^p g(k\boldsymbol{x} - k\boldsymbol{m}) : \boldsymbol{m} \in [\underline{m}, \overline{m}]^p \text{ and } k \in \mathbb{N}\}$, and for any $g \in \mathcal{F}^3$, define the $n$-component bounded finite mixtures of $g$ as the class

$$\mathcal{F}^7_{g,n} = \left\{ f : f(\boldsymbol{x}) = \sum_{i=1}^{n} \pi_i k^p g(k\boldsymbol{x} - k\boldsymbol{m}), \right.$$

$$\left. \boldsymbol{m}_i \in [\underline{m}, \overline{m}]^p, \, k \in \mathbb{N}, \, \pi_i \geq 0, \text{ and } \sum_{i=1}^{n} \pi_i = 1 \right\},$$

where $i \in [n]$ and $-\infty < \underline{m} < \overline{m} < \infty$.

For an arbitrary family of functions $\mathcal{F}$, define the $n$-point convex hull of $\mathcal{F}$ to be

$$\text{Conv}_n(\mathcal{F}) = \left\{ \sum_{i=1}^{n} \pi_i f_i : f_i \in \mathcal{F}, \, \pi_i \geq 0, \text{ and } \sum_{i=1}^{n} \pi_i = 1 \right\},$$

and refer simply to $\text{Conv}_\infty(\mathcal{F}) = \text{Conv}(\mathcal{F})$ as the convex hull. Observe that $\mathcal{F}^7_{g,n} = \text{Conv}_n(\mathcal{F}^6_g)$. By Corollary 1 of [12], we have the fact that

$$\overline{\text{Conv}}(\mathcal{F}^6_g) = \left\{ f : f(\boldsymbol{x}) = \int_{\mathbb{R}^p} k^p g(k\boldsymbol{x} - k\boldsymbol{m}) \, d\Pi_m, \text{ and } \Pi_m \in M_m \right\}$$

March 1, 2018 \hfill DRAFT

is the closure of $\text{Conv}\left(\mathcal{F}_g^6\right)$, where $M_m$ is the sets of all probability measures over $\boldsymbol{m}$. Here, we generically denote the closure of $\text{Conv}\left(\mathcal{F}\right)$ by $\overline{\text{Conv}}\left(\mathcal{F}\right)$. The following result from [20] relates the closure of convex hulls to the $\mathcal{L}_2$-norm.

**Lemma 14** (Barron, 1993, Lemma 1). *If $\bar{f}$ is in $\overline{\text{Conv}}\left(\mathcal{F}\right)$, where $\mathcal{F}$ is a Hilbert space of functions over support $\mathbb{X}$, such that $\|f\|_{\mathbb{X},2}^2 \leq B$ for each $f \in \mathcal{F}$, then for every $n \in \mathbb{N}$, and every $C > B^2 - \|\bar{f}\|_{\mathbb{X},2}^2$, there exists an $f_n \in Conv_n\left(\mathcal{F}\right)$ such that $\|\bar{f} - f_n\|_{\mathbb{X},2}^2 \leq C/n$.*

Thus, from Lemma 14, we know that if $\bar{f} \in \overline{\text{Conv}}\left(\mathcal{F}_g^6\right)$, then there exists an $n$-component finite mixture of density $g$, $f_n \in \mathcal{F}_{g,n}^7$, such that $\|\bar{f} - f_n\|_{\mathbb{K},2}^2 \leq C/n$, where $\mathbb{K}$ is the compact support of both densities and $C > 0$ is a constant that depends on the class $\mathcal{F}_g^6$, which we know to be bounded on $\mathbb{K}$. From Corollary 6 we know that if $f \in \mathcal{F}_{\mathbb{K},\beta}^5 \cap \mathcal{L}_2\left(\mathbb{K}\right)$, then for every $\epsilon > 0$, there exists an $\bar{f} \in \mathcal{F}_g^4$ such that $\|\bar{f} - f\|_{\mathbb{K},2}^2 < \epsilon$. Since $\mathcal{F}_g^4 \subset \overline{\text{Conv}}\left(\mathcal{F}_g^6\right)$, we can set $\bar{f} = f^*$. An application of the triangle inequality yields the following result from [12].

**Theorem 15** (Zeevi and Meir, 1997, Eqn. 27). *If $f \in \mathcal{F}_{\mathbb{K},\beta}^5 \cap \mathcal{L}_2\left(\mathbb{K}\right)$, then for any $\epsilon > 0$ and $g \in \mathcal{F}^3$, there exists an $f_n \in \mathcal{F}_{g,n}^7$ such that $d_\mathbb{K}^{KL}\left(f, f_n\right) \leq \epsilon/\beta + C/\left(n\beta\right)$, for some $C > 0$ and $n \in \mathbb{N}$.*

*Proof:* By the triangle inequality, we have $\|f_n - f\|_{\mathbb{K},2}^2 \leq \|f_n - \bar{f}\|_{\mathbb{K},2}^2 + \|\bar{f} - f\|_{\mathbb{K},2}^2 \leq \epsilon + C/n$. We then apply Lemma 13 to obtain the desired result. ∎

*Remark* 16. The application of Corollary 6 requires the convolution of a compactly supported function with a function over $\mathbb{R}^p$. In general, the convolution of two functions on different supports produces a function with a support that is itself a function of the original supports. That is, if $f$ is supported on $\text{supp}\left(f\right)$ and $g$ is supported on $\text{supp}\left(g\right)$, then the support of $f * g$ is a subset of the closure of the set $\{\boldsymbol{x} + \boldsymbol{y} : \boldsymbol{x} \in \text{supp}\left(f\right), \boldsymbol{y} \in \text{supp}\left(g\right)\}$. In order to mitigate against any problems relating to the algebra of supports, we can allow any compactly supported PDF $f$ to take values outside of its support $\mathbb{K}$ by simply setting $f\left(\boldsymbol{x}\right) = 0$ if $\boldsymbol{x} \notin \mathbb{K}$ and thus implicitly only work with functions over $\mathbb{R}^p$.

*Remark* 17. We note that [12] utilized a slightly different version of Corollary 6 that makes use of the alternative approximate identity $\alpha_k\left(\boldsymbol{x}\right) = k^{-p}g\left(\boldsymbol{x}/k\right)$ with $k^* = 0$. Here, $g$ is taken to be a product kernel of radial basis functions.

An approach for quantifying the error of the quasi-maximum likelihood estimator (quasi-MLE) for finite mixture models, with respect to the Hellinger divergence is then developed by [12] via the theory of [21]. We will instead pursue the bounding of KL errors for the MLE via the directions of [13] and [14].

## IV. MAXIMUM LIKELIHOOD ESTIMATION BOUNDS VIA RESULTS FROM LI AND BARRON (1999)

As alternatives to Lemma 14 and Theorem 15, we can interpret the following results from [13] for finite mixtures of location-scale PDFs over compact supports $\mathbb{K}$.

**Theorem 18** (Li and Barron, 1999, Thm. 1). *If $g \in \mathcal{F}^3$ and $\bar{f} \in \overline{\text{Conv}}\left(\mathcal{F}_g^6\right)$, then there exists an $f_n \in \mathcal{F}_{g,n}^7$ such that $d_{\mathbb{K}}^{KL}\left(\bar{f}, f_n\right) \leq C\gamma/n$, where*

$$C = \int_{\mathbb{K}} \frac{\int_{\mathbb{K}} \left[k^p g\left(k\boldsymbol{x} - k\boldsymbol{m}\right)\right]^2 d\Pi_m}{\int_{\mathbb{K}} k^p g\left(k\boldsymbol{x} - k\boldsymbol{m}\right) d\Pi_m} d\boldsymbol{x}$$

*with DFs $\Pi_m$ over $\mathbb{R}^p$ corresponding to $\bar{f}$, and $\gamma = 4\left(\log\left(3\sqrt{e}\right) + A\right)$ with*

$$A = \sup_{\boldsymbol{m}_1, \boldsymbol{m}_2, \boldsymbol{x}} \log \frac{k^p g\left(k\boldsymbol{x} - k\boldsymbol{m}_1\right)}{k^p g\left(k\boldsymbol{x} - k\boldsymbol{m}_2\right)}.$$

*Remark* 19. Although it is not explicitly mentioned in [13], a condition for the application of Theorem 18 is that $g$ must be such that $A < \infty$ over $\mathbb{K}$. This was alluded to in [14]. This assumption is implicitly made in the sequel.

**Theorem 20** (Li and Barron, 1999, Thm. 2). *For every $\bar{f} \in \overline{\text{Conv}}\left(\mathcal{F}_g^6\right)$ (with corresponding DF $\Pi_m$), if $f \in \mathcal{F}_{\mathbb{K},\beta}^5$ and $g \in \mathcal{F}^3$, then there exists an $f_n \in \mathcal{F}_{g,n}^7$ such that $d_{\mathbb{K}}^{KL}\left(f, f_n\right) \leq d_{\mathbb{K}}^{KL}\left(f, \bar{f}\right) + C\gamma/n$, where $\gamma$ is as defined in Theorem 18, and*

$$C = \int_{\mathbb{K}} \frac{\int_{\mathbb{K}} \left[k^p g\left(k\boldsymbol{x} - k\boldsymbol{m}\right)\right]^2 d\Pi_m}{\left(\int_{\mathbb{K}} k^p g\left(k\boldsymbol{x} - k\boldsymbol{m}\right) d\Pi_m\right)^2} f\left(\boldsymbol{x}\right) d\boldsymbol{x}.$$

By Corollary 6 and Lemma 13, for every $\epsilon > 0$, there exists an $\bar{f} \in \overline{\text{Conv}}\left(\mathcal{F}_g^6\right)$, such that $d_{\mathbb{K}}^{KL}\left(f, \bar{f}\right) < \epsilon/\beta$. We thus have the following outcome.

**Corollary 21.** *If $f \in \mathcal{F}_{\mathbb{K},\beta}^5$ and $g \in \mathcal{F}^3$, then for any $\epsilon > 0$, there exists an $f_n \in \mathcal{F}_{g,n}^7$ such that $d_{\mathbb{K}}^{KL}\left(f, f_n\right) \leq \epsilon/\beta + C\gamma/n$, where $\gamma$ and $C$ are as defined in Theorems 18 and 20.*

Corollary 21 implies that we can approximate a compactly supported PDF to arbitrary degrees of accuracy using finite mixtures of location-scale PDFs of increasing large number of components $n$. Thus far, the results have focused on functional approximation. We now present a KL error bounding result for the MLE.

Let $\boldsymbol{X}_1, ..., \boldsymbol{X}_N$ be $N$ independent and identically distributed (IID) random sample generated from a distribution with density $f \in \mathcal{F}_{\mathbb{K},\beta}^5$. Define the log-likelihood function of an $n$-component mixture of location-scale PDFs $g \in \mathcal{F}^3$ as

$$\ell_{g,n,N}\left(\boldsymbol{\theta}\right) = \sum_{j=1}^{N} \log \left[\sum_{i=1}^{n} \pi_i k^p g\left(k\boldsymbol{X}_i - k\boldsymbol{m}_i\right)\right],$$

where $\boldsymbol{\theta}$ contains $\pi_i$, $k$, and $\boldsymbol{m}_i$ for $i \in [n]$. The MLE can then be defined as

$$\hat{f}_{g,n,N}\left(\boldsymbol{x}\right) = \sum_{i=1}^{n} \hat{\pi}_i k^p g\left(k\boldsymbol{x} - k\hat{\boldsymbol{m}}_i\right),$$

where

$$\hat{\boldsymbol{\theta}}_{n,N} \in \left\{\hat{\boldsymbol{\theta}} : \ell_{g,n,N}\left(\hat{\boldsymbol{\theta}}\right) = \sup \ell_{g,n,N}\left(\boldsymbol{\theta}\right), \text{ satisfying the restrictions of } \mathcal{F}_{g,n}^7\right\}$$

Put the the corresponding estimators of $\pi_i$ and $\boldsymbol{m}_i$ (i.e. $\hat{\pi}_i$, and $\hat{\boldsymbol{m}}_i$) into $\hat{\boldsymbol{\theta}}_{n,N}$. For $B > 0$, if $\mathbb{K}$ is a compact set and the Lipschitz condition

$$\sup_{\boldsymbol{x} \in \mathbb{K}} \left|\log\left[k^p g\left(k\boldsymbol{x} - \boldsymbol{m}_1\right)\right] - \log\left[k^p g\left(k\boldsymbol{x} - \boldsymbol{m}_2\right)\right]\right| \leq B \left\|\boldsymbol{m}_1 - \boldsymbol{m}_2\right\|_1 \quad (2)$$

holds, then the following bound on the expected KL divergence for $\hat{f}_{g,n,N}$ can be adapted from [13].

**Theorem 22** (Li and Barron, 1999, Thm. 3). *Let $g \in \mathcal{F}^3$ and suppose that $\boldsymbol{X}_1, ..., \boldsymbol{X}_N$ is an IID random sample from a distribution with density $f \in \mathcal{F}^5_{\mathbb{K},\beta}$. For every $\epsilon > 0$, if (2) is satisfied and $A = \overline{m} - \underline{m}$, then under the restrictions of $\mathcal{F}^6_g$, there exists a finite $C^* > 0$, such that*

$$\mathbb{E}_f \left[ d^{KL}_{\mathbb{K}} \left( f, \hat{f}_{g,n,N} \right) \right] \leq \frac{\epsilon}{\beta} + \gamma^2 \frac{C^{*2}}{n} + \gamma \frac{2np}{N} \log\left(NABe\right),$$

*$\gamma$ is as in defined in Theorem 18.*

*Proof:* The original theorem provides the inequality

$$\mathbb{E}_f \left[ d^{KL}_{\mathbb{K}} \left( f, \hat{f}_{g,n,N} \right) \right] \leq d^{KL}_{\mathbb{K}} \left(f, \bar{f}^*\right) + \gamma^2 \frac{C^{*2}}{n} + \gamma \frac{2np}{N} \log\left(NABe\right),$$

where $\bar{f}^*$ is the argument that achieves $\inf_{\bar{f} \in \overline{\text{Conv}}(\mathcal{F}^6_g)} d^{KL}_{\mathbb{K}}\left(f, \bar{f}\right)$. By definition $d^{KL}_{\mathbb{K}}\left(f, \bar{f}^*\right) \leq d^{KL}_{\mathbb{K}}\left(f, \bar{f}\right)$, and there exists an $\bar{f} \in \overline{\text{Conv}}\left(\mathcal{F}^6_g\right)$ such that $d^{KL}_{\mathbb{K}}\left(f, \bar{f}\right) < \epsilon/\beta$. Thus select any $\bar{f}$ that satisfies $d^{KL}_{\mathbb{K}}\left(f, \bar{f}\right) < \epsilon/\beta$ and we have the desired result. ∎

*Remark* 23. Since $\epsilon$ can be made as small as we would like, the expected KL divergence between $f$ and the MLE $\hat{f}_{g,n,N}$ can be made arbitrarily small by choosing an increasing sequence of $n$ that grows slower than $N/\log N$. For example, one can take $n = O\left(\log N\right)$. Via some calculus, we obtain the optimal convergence rate by setting $n = O\left(\sqrt{N/\log N}\right)$.

## V. CONCENTRATION INEQUALITIES VIA RESULTS FROM RAKHLIN ET AL. (2005)

We now proceed to utilize the theory of [14] to provide a concentration inequality for the MLE of finite mixtures of location-varying PDFs. Let $\mathcal{N}\left(\Delta, \mathcal{F}, d\right)$ denote the $\Delta$-covering number of the class $\mathcal{F}$, with respect to the distance $d$. That is, $\mathcal{N}\left(\Delta, \mathcal{F}, d\right)$ is the minimum number of $\Delta$-balls that is needed to cover $\mathcal{F}$, where a $\Delta$-ball around $f$ (with centre not necessarily in $\mathcal{F}$) is defined as $\{g : d(f,g) < \delta\}$; see for example [22, Sec. 2.2.2]. Further, define $d_n$ as the empirical distance. That is for functions $f$ and $g$, and realizations $\boldsymbol{x}_1, ..., \boldsymbol{x}_N$ of the random variables $\boldsymbol{X}_1, ..., \boldsymbol{X}_N$, we have $d_n^2(f,g) = N^{-1} \sum_{i=1}^{N} \left[f(\boldsymbol{x}_i) - g(\boldsymbol{x}_i)\right]^2$. The following theorem can be adapted from [14, Thm. 2.1].

**Theorem 24** (Rakhlin et al., 2005, Thm. 2.1). *Let $g \in \mathcal{F}^3$ and suppose that $\boldsymbol{X}_1, ..., \boldsymbol{X}_N$ is an IID random sample from a distribution with PDF $f \in \mathcal{F}^5_{\mathbb{K},\underline{\beta}}$ such that $f(\boldsymbol{x}) < \overline{\beta}$ for all $\boldsymbol{x} \in \mathbb{K}$. If $\hat{f}_{g,n,N}$ is the MLE for an $n$-component finite mixture of $g$ (under the restrictions of $\mathcal{F}^6_g$), then for any $\epsilon > 0$*

$$\begin{aligned}\mathbb{E}_f \left[ d^{KL}_{\mathbb{K}} \left( f, \hat{f}_{g,n,N} \right) \right] &\leq \frac{\epsilon}{\underline{\beta}} + \frac{8\overline{\beta}^2}{n\underline{\beta}^2} \left(2 + \log \frac{\overline{\beta}}{\underline{\beta}}\right) \\ &\quad + \frac{1}{\sqrt{N}} \left( \frac{\overline{\beta}C}{\underline{\beta}^2} \mathbb{E}_f \left( \int_0^{\overline{\beta}} \log^{1/2} \mathcal{N}\left(\Delta, \mathcal{F}^6_g, d_n\right) d\delta \right) + \frac{8\overline{\beta}}{\underline{\beta}} \right) \\ &\quad + \sqrt{\frac{t}{N}} \left( 4\sqrt{2} \log \frac{\overline{\beta}}{\underline{\beta}} \right),\end{aligned}$$

*for some universal constant $C$, with probability at least $1 - \exp(-t)$.*

*Proof:* The original statement of [14, Thm. 2.1] has $d_{\mathbb{K}}^{\text{KL}}\left(f,\bar{f}^{*}\right)$ in place of $\epsilon/\underline{\beta}$. Thus, we obtain the desired result via the same technique as that used in Theorem 22. ∎

*Remark* 25. To make it directly comparable to Theorem 22, one can integrate out the probability statement of Theorem 24 to obtain the inequality in expectation

$$\begin{aligned}\mathbb{E}_f\left[d_{\mathbb{K}}^{\text{KL}}\left(f,\hat{f}_{g,n,N}\right)\right] &\leq \frac{\epsilon}{\underline{\beta}} + \frac{8\overline{\beta}^2}{n\underline{\beta}^2}\left(2+\log\frac{\overline{\beta}}{\underline{\beta}}\right)\\ &\quad + \frac{1}{\sqrt{N}}\left[\frac{\overline{\beta}C}{\underline{\beta}^2}\mathbb{E}_f\left(\int_0^{\overline{\beta}}\log^{1/2}\mathcal{N}\left(\delta,\mathcal{F}_g^6,d_n\right)\mathrm{d}\delta\right)\right]\\ &\quad + \frac{1}{\sqrt{N}}\left(\frac{8\overline{\beta}}{\underline{\beta}} + 4\sqrt{2}\log\frac{\overline{\beta}}{\underline{\beta}}\right).\end{aligned}$$

See the proof of [14, Thm. 2.1] for details. The following corollary specializes the results of Theorem 24 to conform with the conclusion of Theorem 22.

**Corollary 26** (Rakhlin et al., 2005, Cor. 2.2). *Let $g \in \mathcal{F}^3$ and suppose that $\boldsymbol{X}_1,...,\boldsymbol{X}_N$ is an IID random sample from a distribution with density $f \in \mathcal{F}_{\mathbb{K},\underline{\beta}}^5$ such that $f(\boldsymbol{x}) < \overline{\beta}$ for all $\boldsymbol{x} \in \mathbb{K}$. For every $\epsilon > 0$, if (2) is satisfied and $A = \overline{m} - \underline{m}$, under the restrictions of $\mathcal{F}_g^6$,*

$$\mathbb{E}_f\left[d_{\mathbb{K}}^{KL}\left(f,\hat{f}_{g,n,N}\right)\right] \leq \frac{\epsilon}{\underline{\beta}} + \frac{C_1}{n} + \frac{C_2}{\sqrt{N}},$$

*where $C_1$ and $C_2$ are constants that depend on $\underline{\beta}$, $\overline{\beta}$, $A$, $B$, $C$, and $p$. Here, $C$ is the same universal constant as in Theorem 24.*

*Remark* 27. Corollary 26 directly improves upon the result of Theorem 22 by allowing $n$ and $N$ to increase independently of one another and still be able to achieve an arbitrarily small bound on the expected KL divergence of the MLE for finite mixtures of location-scale PDFs, under the same hypothesis. The corollary implies that the optimal choice for the number of components is to set $n = O\left(\sqrt{N}\right)$.